\documentclass[aps,prb,twocolumn,amsmath,amssymb,superscriptaddress,floatfix]{revtex4}
\usepackage{graphicx}
\usepackage{graphics}
\usepackage{bm}
\usepackage[usenames]{color}
\usepackage{colordvi}
\usepackage{units}
\usepackage{bbm}
\usepackage{booktabs}
\usepackage{array}
\usepackage{placeins}
\usepackage{epsfig}
\usepackage{lscape}
\usepackage{changes}
\usepackage{braket}
\usepackage{tabularx}
\newcolumntype{L}[1]{>{\raggedright\arraybackslash}p{#1}} 
\newcolumntype{C}[1]{>{\centering\arraybackslash}p{#1}} 
\newcolumntype{R}[1]{>{\raggedleft\arraybackslash}p{#1}} 
\usepackage[colorlinks=true,linkcolor=blue,citecolor=blue,urlcolor=blue]{hyperref}
\newcommand{\be}{\begin{equation}}
\newcommand{\ee}{\end{equation}}
\newcommand{\beqn}{\begin{eqnarray}}
\newcommand{\eeqn}{\end{eqnarray}}

\definecolor{mymagenta}{rgb}{1.0,0.0,1.0}
\definecolor{mycyan}{rgb}{0.0,1.0,1.0}
\definecolor{myyellow}{rgb}{1.0,1.0,0.0}
\definecolor{myorange}{rgb}{1.0,0.27,0.0}

\definecolor{dark-gray}{HTML}{a0a0a0}
\definecolor{dark-red}{HTML}{8b0000}
\definecolor{dark-green}{HTML}{006400}
\definecolor{dark-blue}{HTML}{00008b}
\definecolor{gold}{rgb}{1.0,0.84,0.0}
\definecolor{dark-turquoise}{HTML}{00ced1}

\bibstyle{apsrev.bib}

\begin{document}

\title{Mixed-order transition in the antiferromagnetic quantum Ising chain in a field}
\author{P\'eter Lajk\'o}
\email{peter.lajko@ku.edu.kw}
\affiliation{Department of Physics, Kuwait University, P.O. Box 5969, Safat 13060, Kuwait}
\author{Ferenc Igl{\'o}i}
\email{igloi.ferenc@wigner.hu}
\affiliation{Wigner Research Centre for Physics, Institute for Solid State Physics and Optics, H-1525 Budapest, P.O. Box 49, Hungary}
\affiliation{Institute of Theoretical Physics, Szeged University, H-6720 Szeged, Hungary}
\date{\today}

\begin{abstract}
The antiferromagnetic quantum Ising chain has a quantum critical point which belongs to the universality class of the transverse Ising model (TIM). When a longitudinal field ($h$) is switched on, the phase transition is preserved, which turns to first-order for $h/\Gamma \to \infty$, $\Gamma$ being the strength of the transverse field. Here we will re-examine the critical properties along the phase transition line. During a quantum block renormalization group calculation, the TIM fixed point for $h/\Gamma>0$ is found to be unstable. Using DMRG techniques, we calculated the entanglement entropy  and the spin-spin correlation function, both of which signaled a divergent correlation length at the transition point with the TIM exponents. At the same time, the bulk correlation function has a jump and the end-to-end correlation function has a discontinuous derivative at the transition point. Consequently for finite $h/\Gamma$ the transition is of mixed-order.
\end{abstract}


\maketitle

\section{Introduction}
\label{sec:intr}
Quantum phase transitions are among the fundamental problems of modern physics, the properties of which are studied in different disciplines: solid state physics, quantum field theory, quantum information, and statistical mechanics\cite{sachdev}. Quantum phase transitions take place at $T = 0$ temperature and these are indicated by singularities in the ground-state expectational values of some observables by varying a control parameter, such as the strength of a transverse field. There are several model systems which has been experimentally realised by ultra-cold atoms in an optical lattice\cite{lewenstein}. One of such basic problems is the antiferromagnetic Ising chain in a mixed transverse and longitudinal field. The Hamiltonian of the model is defined as:
\begin{align}
\begin{split}
\hat{H}&=\sum_{i=1}^L J \sigma_{i}^{z} \sigma_{i+1}^{z}\\
&-\sum_{i=1}^L \Gamma \sigma^x_{i} -h\sum_{i=1}^L \sigma^z_{i}\;.
\label{Hamilton}
\end{split}
\end{align}
in terms of the $\sigma_{i}^{x,z}$ Pauli matrices at site $i$. In the experimental setup the longitudinal fields are not too strong\cite{simon}.

Theoretically, the quantum phase transition in this model has been studied by different methods: finite-size exact diagonalization by the L\'anczos method\cite{sen}, density matrix renormalisation (DMRG)\cite{ovchinnikov,lajko}, quantum Monte Carlo simulations\cite{Yu_Cheng} and the fidelity susceptibility method\cite{bonfim}. The limiting case, $h/J \to 2$ and $\Gamma \to 0$ is studied in Ref.\cite{hard_rods} and a generalisation with an $m$-spin product term, i.e. replacing $\sigma_{i}^{z} \sigma_{i+1}^{z}$ by $\prod_{j=0}^{m-1} \sigma_{i+j}^z$  is studied in Refs.\cite{penson,rods_MC1,rods_MC2}. Recently, the model with quenched disorder, in which $J_i>0$ and $\Gamma_i$ are i.i.d. random numbers, but the $h_i=h$ are non-random, has also been studied\cite{Yu_Cheng,lajko}.

\begin{figure}[h!]
\begin{center}
\includegraphics[width=8.6cm,angle=0]{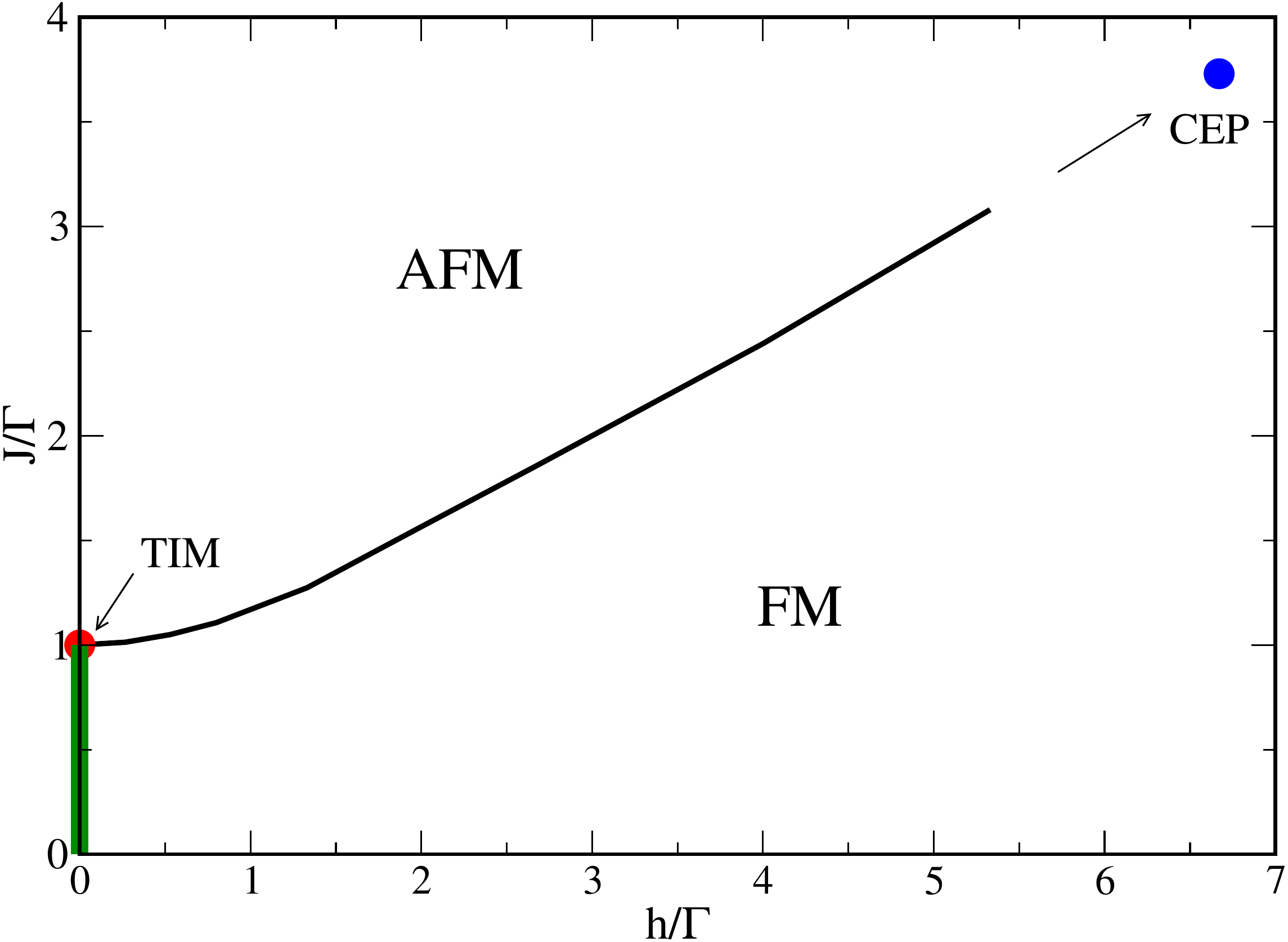}
\end{center}
\caption{\label{fig:clean_phase}(Color online) The zero-temperature phase diagram of the Ising chain with antiferromagnetic coupling, $J$,
in transverse ($\Gamma$) and longitudinal ($h$) magnetic fields calculated by the DMRG method. 
In the absence of longitudinal field, $h=0$, the transition between a quantum antiferromagnetic (AFM) phase and a quantum paramagnetic (PM) phase (indicated by thick green line) is controlled by the fixed point of the transverse Ising model (TIM) at $(J/\Gamma=1,\,h/\Gamma=0)$. For finite value of $h>0$ the AFM phase survives and at the other side of the phase boundary there is a quantum ferromagnetic (FM) phase. In the classical limit, $\Gamma=0$, thus $J/\Gamma \to \infty$ and $h/\Gamma \to \infty$ there is a classical end-point (CEP) at $(h/J=2)$ having a first-order transition.  
}
\vskip-3mm
\end{figure}

The phase-diagram of the clean system in Eq.(\ref{Hamilton}) is calculated through the extrapolated position of the maximum of the entanglement entropy, as described in Sec.\ref{sec:entropy_max}. It is shown in Fig.\ref{fig:clean_phase} in terms of $J/\Gamma$ vs. $h/\Gamma$. This has two exactly solved points, which are denoted by TIM and CEP, respectively. In the absence of longitudinal field, $h=0$, there is an antiferromagnetic (AFM) ordered phase, for $J/\Gamma>1$, which is separated from the paramagnetic (PM) phase by a continuous phase transition point, at $J_c/\Gamma=1$ which belongs to the universality class of the transverse Ising model (TIM)\cite{pfeuty}. 
In the vicinity of the transition point the excitation energy (inverse correlation length) $\epsilon \sim 1/\xi$ vanishes as:
\be
\epsilon \sim J-J_c,\quad J \ge J_c
\label{TIC_gap}
\ee
thus the gap (correlation length) exponent is $\nu=1$. The AFM order is characterised by the spin-spin correlation function:
\be
C(r)=(-1)^r\langle \sigma_i^z \sigma_{i+r}^z \rangle\;,
\label{corr_func}
\ee
which is translational invariant for periodic chains. At the phase-transition point $C(r)$ has a power-law decay:
\be
C(r) \sim r^{-\eta}\;,
\label{TIC_corr}
\ee
with $\eta=1/4$.
The second exactly solved point is at $J/\Gamma \to \infty$ and $h/\Gamma \to \infty$, in which quantum fluctuations are negligible. Here the phase transition takes place at the classical end-point (CEP), at $h/J=2$ between an AFM phase and a ferromagnetic (FM) phase and the transition is of first order. At the CEP the ground state is infinitely degenerate, the entropy per site is finite\cite{domb}.

Our interest in this work is the behaviour of the system between the two exactly solved points, for $0<h/\Gamma<\infty$. Here the phase-diagram contains an AFM ordered phase for strong enough couplings, $J/\Gamma > J_c/\Gamma$. At the other side of the phase-transition line, for $J/\Gamma < J_c/\Gamma$ there is a ferromagnetic (FM) order in the system, which is maintained by the longitudinal field. This is evident at $J=0$, in which case $C(r)=(-1)^r[(\Gamma/h)^2+1]^{-1}$. In previous studies\cite{ovchinnikov,hard_rods} the excitation spectrum of the system has been investigated and a vanishing gap is found at the phase-transition line, $J=J_c$, signalling a divergent correlation length. According to precise numerical results the gap vanishes in the same way as for the TIM in Eq.(\ref{TIC_gap}). Due to this result it is generally believed that the phase-transition for $0<h/\Gamma <\infty$ belongs to the TIM universality class.

With this claim, however, there are still some unclear points. It was not investigated how the correlation function behaves when passing the phase transition line from the AFM-ordered phase. At $h/\Gamma=0$ it stays zero in the complete PM phase, but for $h/\Gamma>0$ it should become ferromagnetic at some point. It is of interest how this transition from AFM to FM orders occurs. A further question how this transition is connected with the first-order one at the CEP in the transition line. It is also interesting, if this transition is symmetric or not, at least in terms of the critical exponents at the two sides of the transition. Another open question concerns the scaling behavior of the entanglement entropy in the vicinity of the phase transition point. 

Our aim in this article is to examine and clarify the above issues. First, we studied the stability of the TIM fixed point in relation to switching on the longitudinal field. For this we used an approximate quantum block renormalisation group (RG) method and let vary the size of the block. More precise numerical results are obtained from a DMRG calculation. We investigated the scaling properties of the entanglement entropy in the phase-transition region and studied the divergency of the related correlation length. In extensive numerical work the spin-spin correlation function is calculated and its properties are analysed, both for bulk spins and for end-to-end correlations.

The rest of the paper is organised in the following way. In Sec.\ref{sec:RG} we introduce and apply the quantum block RG method to test the stability of the fixed point of the TIM. The DMRG results are presented in Sec.\ref{sec:dmrg}, first about the scaling behaviour of the entanglement entropy in Sec.\ref{sec:entanglement_entropy} and afterwards about the correlation function in Sec.\ref{sec:corr}. Our paper is closed with a discussion in Sec.\ref{sec:disc}.

\section{Quantum RG treatment}
\label{sec:RG}

Renormalisation of quantum systems started with the celebrated paper by Wilson about the Kondo problem\cite{wilson}. For lattice models a real space version, the so called block transformation method has been introduced\cite{drell} and applied for a set of (mainly one-dimensional) problems\cite{review}. In this method the system is divided into blocks and the parameters of the Hamiltonian are separated to intra- and inter-block terms. Solving the block Hamiltonian exactly by numerical methods the lowest states are retained and identified as states of the block-spin variable, while the renormalised values of the inter-block parameters are obtained by (first-order) perturbation calculation. Due to the used approximations the block RG method generally provides rather poor results with the smallest block-size ($n=2$), however using larger blocks the results have been improved. This has been shown for the TIM\cite{uzelac,igloi} among others, i.e. with the Hamiltonian in Eq.(\ref{Hamilton}) in the absence of a longitudinal field, $h=0$.

\begin{figure}[h!]
\begin{center}
\includegraphics[width=8.6cm,angle=0]{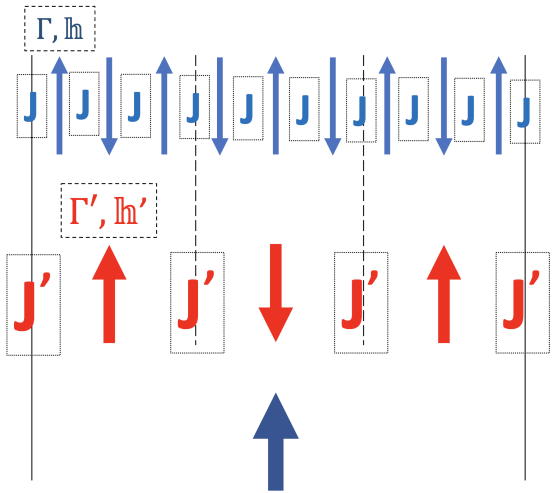}
\end{center}
\caption{\label{fig:RG}(Color online) Illustration of the block renormalization for $n=3$.
}
\end{figure}

Here in our treatment we include the longitudinal field, both in the Hamiltonian and in the RG transformations. In particular we are interested in the stability of the TIM fixed-point by switching on the longitudinal field. Since at the CEP there are no quantum fluctuations our quantum RG can't describe the behaviour of the system around that classical point. In this procedure the length of the blocks, $n$, is an \textit{odd number}, in order to preserve the antiferromagnetic order in the system. During renormalization the block of spins is replaced by a single renormalised Ising spin variable, which is illustrated in Fig.\ref{fig:RG}. For this the block Hamiltonian is solved exactly and from its spectrum the two lowest eigenstates are retained. The renormalised values of the parameters are obtained from the condition that the matrix-elements are equal in the original and in the renormalised basis.

In the actual calculation first we perform a gauge transformation: $\sigma_{i}^{z} \to (-1)^i \sigma^z_{i}$ and make a rotation $x \leftrightarrow z$, thus the intra-block Hamiltonian of $n$ spins is written as:
\begin{align}
\begin{split}
\hat{H}_B&=-\sum_{i=1}^{n-1} J \sigma_{i}^{x} \sigma_{i+1}^{x}\\
&-\sum_{i=1}^n \Gamma \sigma^z_{i} -h\sum_{i=1}^n (-1)^i \sigma^x_{i}\;.
\label{Hamilton_B}
\end{split}
\end{align}
We solve the ground state ($|0\rangle$) and the first excited state ($|1\rangle$) of $\hat{H}_B$ with energies $\epsilon_0$ and $\epsilon_1$ respectively. We also calculate the matrix-elements: $m_x(i)=\langle 0 |\sigma^x_i|1\rangle$ and $m_z(i)=\langle 0 |\sigma^z_i|0\rangle$, which are called as $x$ and $z$ magnetisations, respectively. For the block the total magnetisations are $M_x=\sum_{i=1}^n m_x(i)$ and $M_z=\sum_{i=1}^n m_z(i)$.

The renormalised value of the coupling follows from:
\be
J'=J m_x(1) m_x(n)\;.
\label{J'}
\ee
The renormalised values of the transverse ($\Gamma'$) and the longitudinal fields ($h'$) are related with the value of the gap:
\be
\sqrt{(\Gamma')^2+(h')^2}=\frac{\epsilon_1-\epsilon_0}{2}\;.
\label{Gamma'h'}
\ee
In addition we require that the ratio of the total $x$ and $z$ magnetisations, ${M_x}/{M_z}$, remains the same after the transformation. This leads to the relation:
\be
\frac{\Gamma'}{h'}=\frac{M_x}{M_z}\;.
\label{Gamma'h'_ratio}
\ee

We have iterated the RG equations in Eqs.(\ref{J'}),(\ref{Gamma'h'}) and (\ref{Gamma'h'_ratio}), which has two trivial fixed points. The one at ($J/\Gamma=0,h/\Gamma=0$) controls the disordered phase, while the other at ($J/\Gamma=\infty,h/\Gamma=0$) is for the ordered phase. The transition between these phases are controlled by two non-trivial fixed points. The first, denoted by $FP1$ is located at $(J/\Gamma)_1>0$ and $(h/\Gamma)_1=0$ is the TIM fixed poin. Without longitudinal field ($h=0$) its properties have been studied earlier in details\cite{uzelac,igloi}. Here we are interested in its stability with respect to switching on the longitudinal field. The second non-trivial fixed pont, $FP2$ is located at
$(J/\Gamma)_2>0$ and $(h/\Gamma)_2>0$. At the non-trivial fixed points we have calculated the eigenvalues of the linearised transformation, $\lambda_1>\lambda_2$ from which the scaling exponents follows as $y_{1,2}=\frac{\ln \lambda_{1,2}}{\ln n}$. As usual, for $y>0$ ($y<0$) the fixed point is repulsive (attractive). For $FP2$ we found $\lambda_2<0$, which means rotation of the trajectory in the vicinity of the fixed point. This behaviour is related to the quantum nature of the RG, which cannot describe a first-order transition between classical states at CEP. We have also calculated the dynamical exponent, $z$, from the scaling of the energy at the fixed point: $\Gamma'=n^{-z}\Gamma$. For different values of the size of the block, $n=3,5,7,9,11$ and $13$ the calculated critical parameters are collected in Tables \ref{Tab01} and \ref{Tab02}.

\begin{table}
\renewcommand{\arraystretch}{1.5}
\begin{tabular}{ C{0.6cm} | C{1.4cm} | C{1.1cm} | C{1.4cm}| C{1.4cm} | C{1.4cm} }
 $n$ & $(J/\Gamma)_1$ & $(h/\Gamma)_1$ & $y_1$ & $y_2$ & $z$ \tabularnewline
\hline
$3$ & $0.8660$ & $0.$ & $0.763$ &  $0.011$ & $0.631$\tabularnewline
\hline
$5$ & $0.9262$ & $0.$ & $0.833$ &  $0.119$ & $0.704$\tabularnewline
\hline
$7$ & $0.9496$ & $0.$ & $0.863$ &  $0.169$ & $0.741$\tabularnewline
\hline
$9$ & $0.9619$ & $0.$ & $0.879$ &  $0.198$ & $0.763$\tabularnewline
\hline
$11$ & $0.9694$ & $0.$ & $0.890$ &  $0.217$ & $0.779$\tabularnewline
\hline
$13$ & $0.9745$ & $0.$ & $0.898$ &  $0.230$ & $0.791$\tabularnewline
\hline
$\infty$ & $1.$ & $0.$ & $1.$ &  $ - $ & $1.$\tabularnewline
\end{tabular}
\caption{Critical parameters at the $FP1$ fixed point calculated with different size of the block, $n$.
\label{Tab01}}
\end{table}
Regarding $FP1$ the critical parameters (except $y_2$) agree with the previous calculations in Refs.\cite{uzelac,igloi}. The exact results, which are obtained for $n \to \infty$ and presented in the last row of Table \ref{Tab01} are approached logarithmically, having a correction $\sim 1/\ln n$. The new feature of this calculation is the second exponent, $y_2$, due to the presence of the longitudinal field. Since $y_2>0$, according to our RG approach $h>0$ is a relevant perturbation, thus the critical properties of the complete model should be different to the TIM. Having a look at the block-size dependence of $y_2$ it smoothly increases with $n$, for small sizes this grows approximately as $\sim 0.09 \ln n$.
\begin{table}
\renewcommand{\arraystretch}{1.5}
\begin{tabular}{ C{0.6cm} | C{1.4cm} | C{1.1cm} | C{1.4cm}| C{1.4cm} | C{1.4cm} }
 $n$ & $(J/\Gamma)_2$ & $(h/\Gamma)_2$ & $y_1$ & $\lambda_2$ & $z$ \tabularnewline
\hline
$3$ & $1.1052$ & $0.8998$ & $1.052$ &  $0.114$ & $0.640$\tabularnewline
\hline
$5$ & $1.0079$ & $0.6830$ & $0.955$ &  $-0.042$ & $0.657$\tabularnewline
\hline
$7$ & $0.9791$ & $0.5852$ & $0.939$ &  $-0.140$ & $0.667$\tabularnewline
\hline
$9$ & $0.9672$ & $0.5243$ & $0.916$ &  $-0.216$ & $0.673$\tabularnewline
\hline
$11$ & $0.9617$ & $0.4808$ & $0.899$ &  $-0.279$ & $0.677$\tabularnewline
\hline
$13$ & $0.9591$ & $0.4473$ & $0.887$ &  $-0.333$ & $0.681$\tabularnewline
\end{tabular}
\caption{Critical parameters at the $FP2$ fixed point calculated with different size of the block, $n$.
\label{Tab02}}
\end{table}

As seen in Table \ref{Tab02} for the position of $FP2$, that $(J/\Gamma)_2 \approx (J/\Gamma)_1$, while $(h/\Gamma)_2$ decreases with $n$. For small block-sizes we have an approximate relation $(h/\Gamma)_2 \sim 1.15/\ln n$. The leading critical index, $y_1$ slowly decreases with $n$ and it is approximately the same at both fixed points. We can conclude the results of this approximate investigation, that the critical behaviour of the model for $h>0$ is probably not controlled by the TIM fixed point, but some critical indices are still very close to the quantum Ising values. We expect more accurate information from DMRG studies.

\section{DMRG results}
\label{sec:dmrg}

The complete phase-diagram of the model has been studied numerically by the DMRG method. In this investigation we used finite samples of length up to $L=1024$, their ground state and the first few excited states are calculated. Here the original version of the finite system DMRG scheme was utilized with periodic and open boundary conditions\cite{dmrg1,dmrg2}. For the
correlations it was systematically and carefully checked that their values are vastly independent of the basis size
$m$ within the error bars. The accuracy of the ground-state energy calculations was in the range of $10^{-6} \sim 10^{-8}$ and this was in full agreement with the truncation error, the largest basis size being $m = 140 \sim 260$ for the different systems. 
 
In general we have fixed the value of the transverse field to $\Gamma=1.5$ and used different values of the longitudinal field: $h=0.4, 0.8, 2.0, 10.0, 20.0$ and $40.0$. By varying the strength of the coupling we have studied the behaviour of the correlation function, ${C(L/2-1)}=-\langle \sigma_{1}^{z} \sigma_{L/2-1}^{z}\rangle$
as well as the entanglement entropy, ${\cal S}$ calculated between the two halves of the chain. Particular attention is made to the properties of the system in the vicinity of the phase transition point.

\subsection{Entanglement entropy}
\label{sec:entanglement_entropy}

If we divide the chain into two parts of lengths $\ell$ and $L-\ell$, then the entanglement between the two parts is quantified by the von Neumann entropy:
\be
{\cal S}_{\ell}=-\mathbf{Tr}(\rho_{\ell} \ln \rho_{\ell})=-\mathbf{Tr}(\rho_{L-\ell} \ln \rho_{L-\ell})={\cal S}_{L-\ell}\;.
\label{entropy}
\ee
Here $\rho_{\ell}=\mathbf{Tr}_{L-\ell} \rho$ and $\rho_{L-\ell}=\mathbf{Tr}_{\ell} \rho$ is the reduced density matrix with $\rho=|\phi\rangle\langle \phi |$ and $|\phi\rangle$ being a pure state of the closed quantum system. In our calculation we have $\ell=L/2$ and periodic boundary conditions are used, when the two parts have $b=2$ contact points. For brevity in the following we use the notation: ${\cal S}_{\ell=L/2}={\cal S}$.

\subsubsection{Finite-size scaling of the maximum}
\label{sec:entropy_max}

\begin{figure}[h!]
\begin{center}
\includegraphics[width=8.6cm]{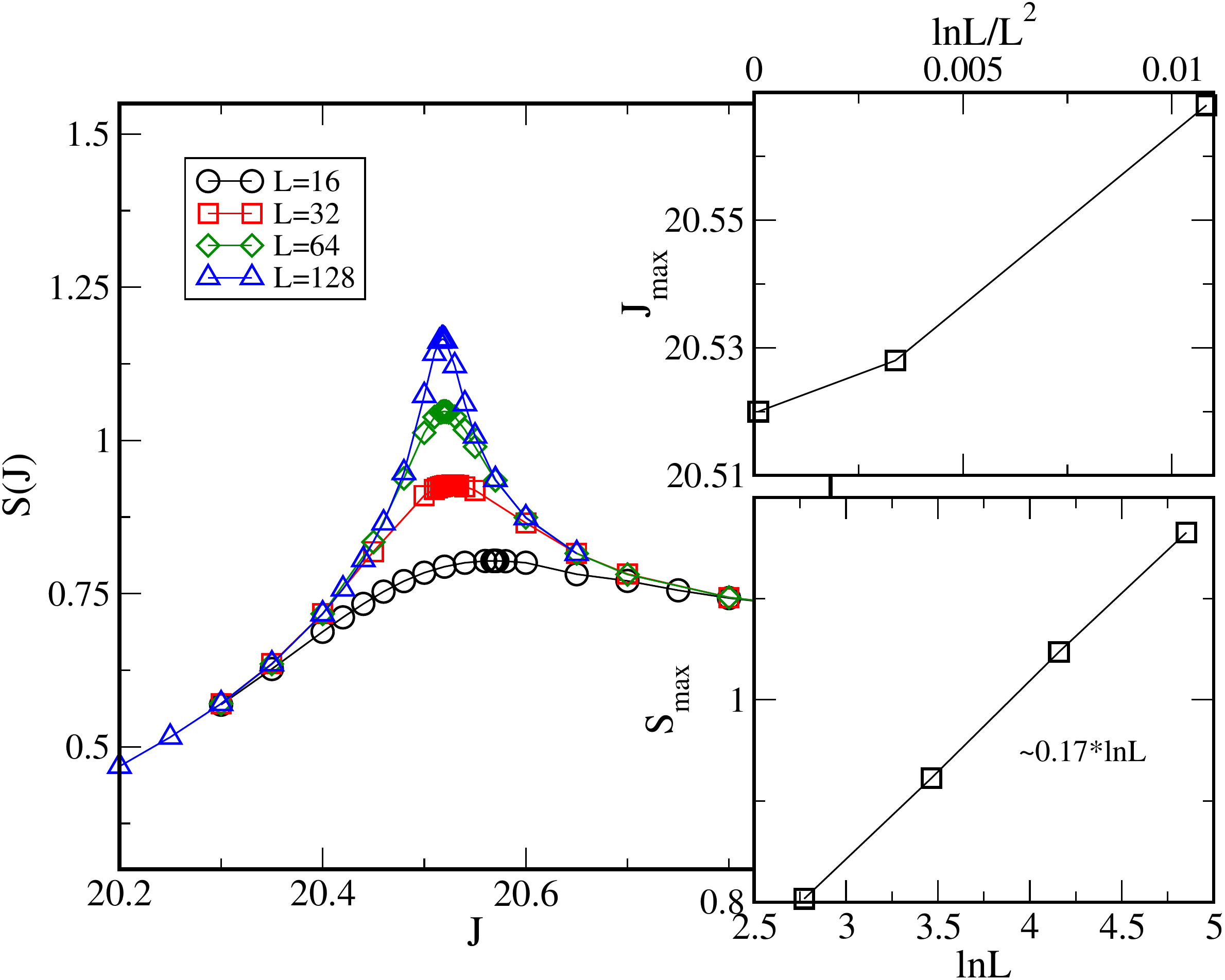}
\end{center}
\caption{\label{fig:entropy}(Color online) The $J$ dependence of the entanglement entropy at $h=40.0$ and $\Gamma=1.5$ for different lengths of periodic chains. The position of the maximum is shifted by $\sim \ln L/L^2$ from the phase transition point, see in the upper inset. The value at the maximum scales as $\approx 0.17 \ln L$, see in the lower inset.
}
\end{figure}

The $J$-dependence of the entanglement entropy at $h=40.0$ is shown in Fig.\ref{fig:entropy} for different lengths of the chain. ${\cal S}(J)$ has a maximum, the position of which $J_{\rm max}(L)$ can be used as a finite-size, pseudo phase-transition point\cite{igloi07}. It approaches the true transition point, $J_c$ as $J_{\rm max}(L)-J_c \sim \ln L/L^2$. The logarithmic correction term has been detected at $h=0$ and it seems to be present for $h>0$ as well, which is illustrated in the upper inset of Fig.\ref{fig:entropy}. Using this correction term the phase-transition point can be accurately determined. We obtained the following shift of the critical point close to $h=0$ as $J_c/\Gamma-1 \approx 0.19 (h/\Gamma)^2$, in agreement with Ref.\cite{Yu_Cheng}. The value of the maximum, ${\cal S}_{\rm max}(L)$, which is very close to ${\cal S}(J_c,L)$ is found to scale logarithmically:
\be
{\cal S}_{\rm max}(L) \simeq b \frac{c}{6} \ln L\;,
\label{entropy_L}
\ee
which is illustrated in the lower inset of Fig.\ref{fig:entropy}.
At a second-order transition point for conformally invariant systems $c$ is the central charge of the Virasoro algebra\cite{holzhey,korepin,calabrese}. In Fig.\ref{fig:entropy} we obtained an estimate: $bc/6 \approx 0.17$. Repeating the same analysis at $h=0.4$ a more accurate estimate is found: $bc/6 \approx 0.167$, whereas at $h=0$ the exact value is $bc/6 = 1/6$, thus $c=1/2$. Our numerical results about the entanglement entropy indicate that its finite-size scaling is compatible with the central charge of the Ising model not only at $h=0$, but also for $h>0$.

\subsubsection{The correlation length}

\begin{figure}[h!]
\begin{center}
\includegraphics[width=8.6cm]{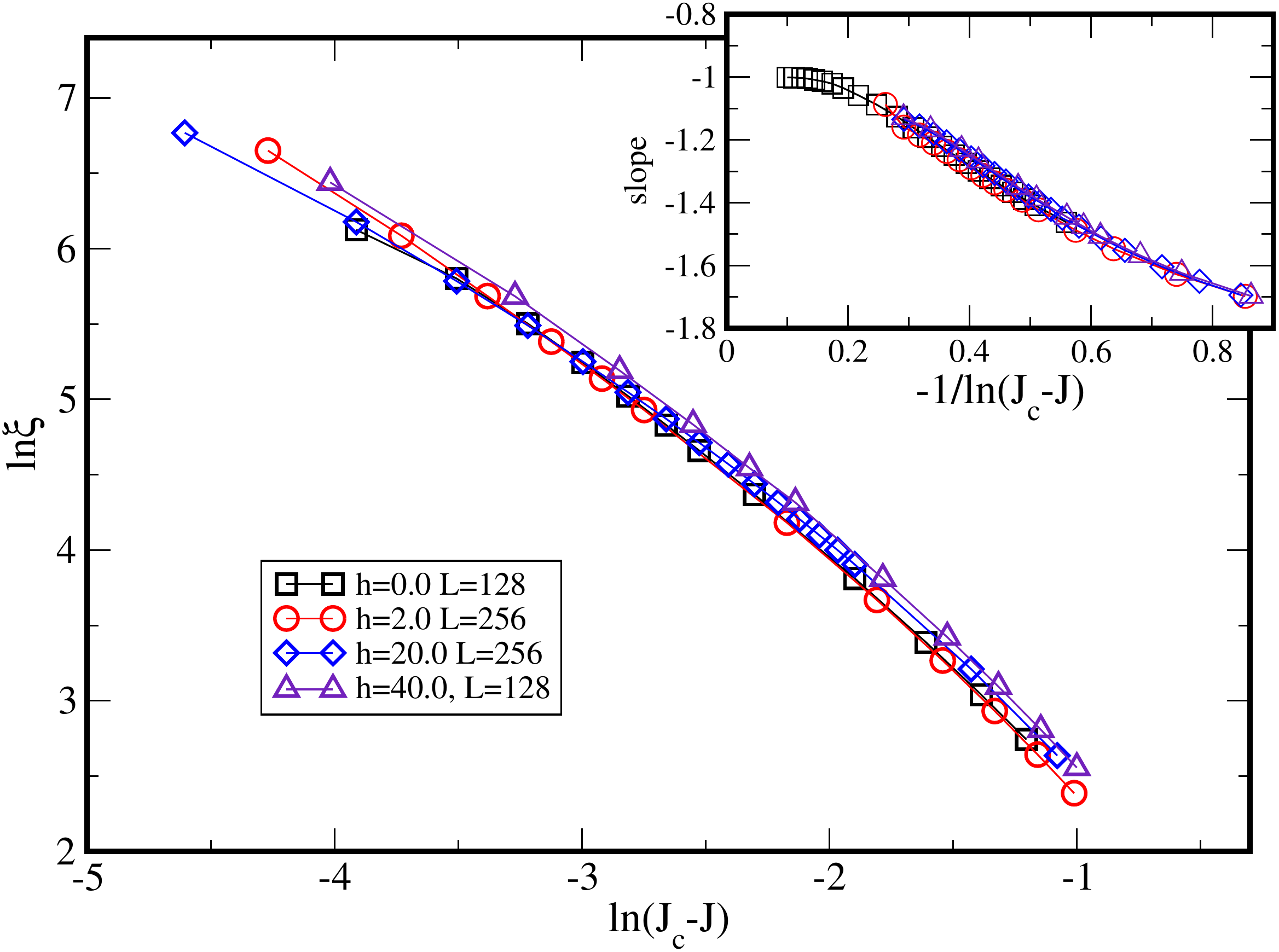}
\end{center}
\caption{\label{fig:corr_length}(Color online) The correlation length, extracted from the entropy through Eq.(\ref{entropy_xi}) as a function of the difference from the transition point in the disordered phase in log-log scale.
In the inset the local slope of the curves, approaching the critical exponent $-\nu$ is shown. For the case $h=0$ the slope is taken from the analytical result in Ref.\cite{igloi08} .
}
\end{figure}

The correlation length, $\xi$ is extracted from the entanglement entropy, using the property, that in the vicinity of the critical point ${\cal S}$ behaves as
\be
{\cal S} \simeq b \frac{c}{6} \ln \xi\;.
\label{entropy_xi}
\ee
In the disordered phase, $J < J_c$ fixing $c=1/2$ we show in Fig.\ref{fig:corr_length} $\ln \xi$ as a function of $\ln(J_c-J)$ for different values of $h$. It is seen that the curves for different $h$ are very close to each other and therefore the same type of asymptotic scaling is expected, as for $h=0$. We have also calculated the slope of the curves, shown in the inset of Fig.\ref{fig:corr_length}, which is expected to approach $-\nu=-1$, as known analytically at $h=0$. We conclude that the correlation length critical exponent in the disordered phase for all values of $h>0$ is compatible with the Ising value $\nu=1$.
Expecting that this relation holds for large $h$, where $h/J_c \approx 2$ the asymptotic region of the numerical points is in accordance with the relation: 
\be
\xi(h) \sim \frac{\Gamma}{J_c-J} \sim \left(\frac{J_c}{J_c-J}\right) \frac{\Gamma}{J_c} \sim \Delta^{-1} \frac{\Gamma}{h_c}\sim \frac{\Gamma}{h-h_c}\;.
\ee
with $\Delta=(J_c-J)/J_c=(h-h_c)/h_c$ being the reduced control-parameter at the transition point. Now taking $\Gamma \sim \Gamma_c$ we notice that the correlation length at the CEP tends to a finite value, since $\Gamma_c \sim (2-h)$. In this way we have a smooth behaviour of $\xi$ around the CEP.

\subsection{Correlation function}
\label{sec:corr}

The order in the ground state is characterised by the correlation function in Eq.(\ref{corr_func}). In general, we consider correlations between distant points, fix $r=L/2-1$ and for brevity we use the notation $C(L/2-1)=C$, which is the function of $J,\Gamma$ and $h$.

\begin{figure}[h!]
\begin{center}
\includegraphics[width=8.6cm]{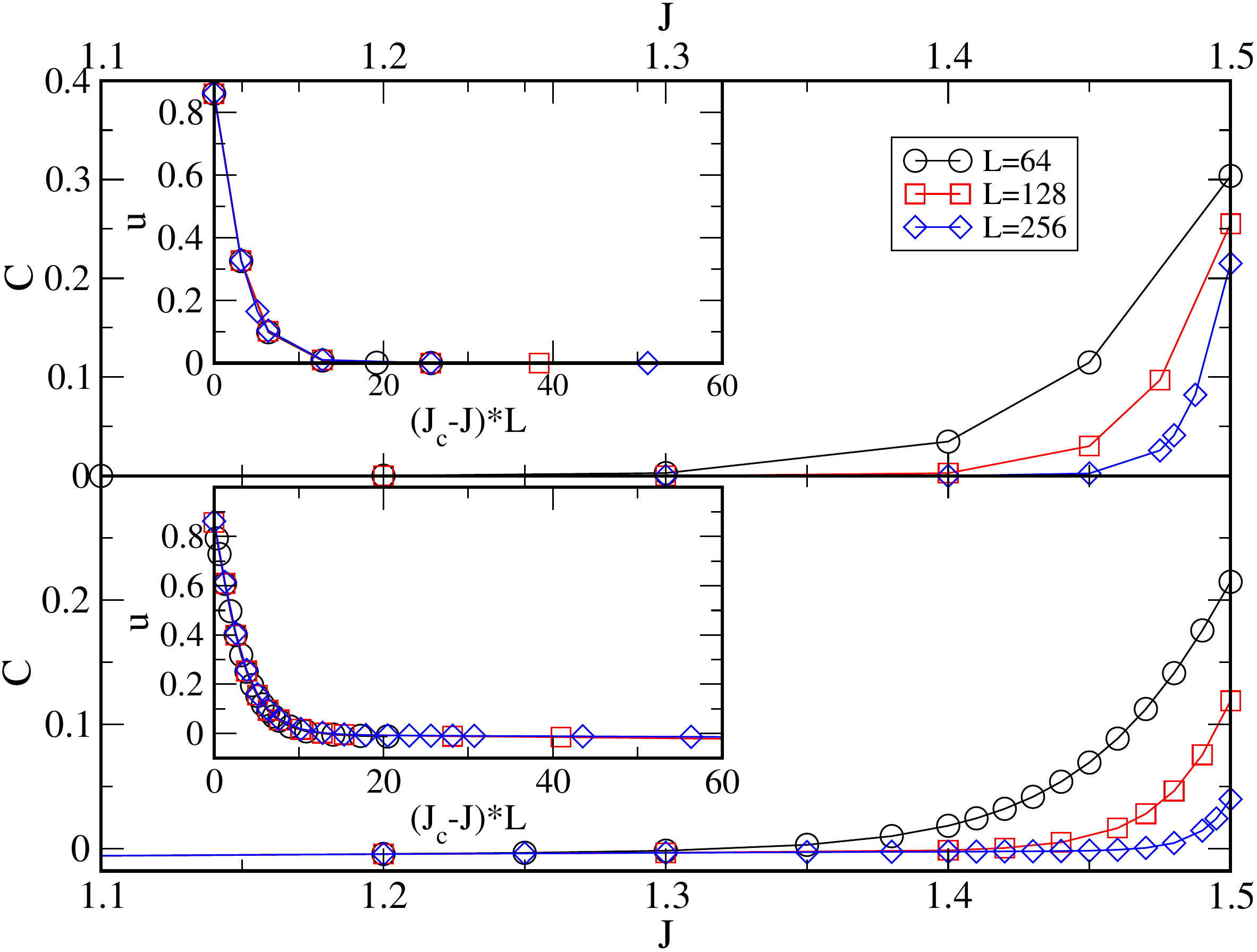}
\end{center}
\caption{\label{fig:corr_func1}(Color online) Correlation function at $\Gamma=1.5$, $h=0.$ (upper panel) and $h=0.4$ (lower panel) as a function of the coupling for different lengths. In the insets the scaling plots $u=C(J,L)L^{1/4}$ vs. $(J_c-J)L$ are shown.}
\end{figure}

\begin{figure}[h!]
\begin{center}
\includegraphics[width=8.6cm]{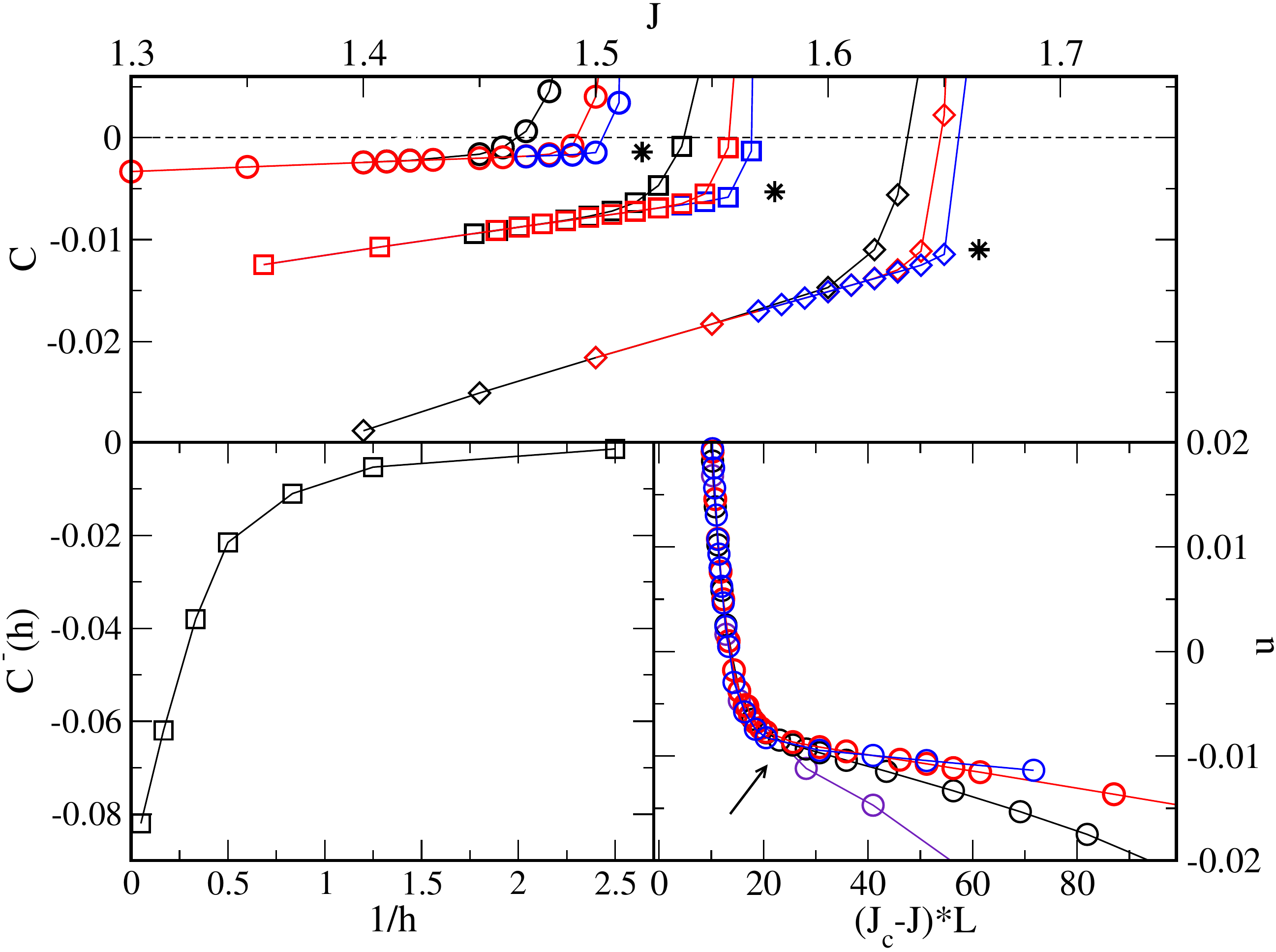}
\end{center}
\caption{\label{fig:corr_func2}(Color online) Correlation function at $h=0.4$ (denoted by $\circ$), $h=0.8$ (denoted by $\square$) and $1.2$ (denoted by $\diamond$) as a function of the antiferromagnetic coupling, $J>0$, close to the transition point, calculated by DMRG at different lengths. $L=256$ $\circ$, $L=512$ \textcolor{red}{$\circ$}, $L=1024$ \textcolor{blue}{$\circ$}. The limiting value of the correlation function, as defined in Eq.(\ref{C_limit}) is indicated by an $\ast$. The limiting values as a function of $1/h$ are plotted in the left inset. In the right inset a scaling plot of $u=C(J,L)L^{1/4}$ vs. $(J_c-J)L$ is shown for $h=0.4$. The point indicated by an arrow is the critical end-point, which separates the scaling and non-scaling regions.
}
\end{figure}

If there is no interaction, $J=0$, then the correlation function is ferromagnetic, $C(J=0)=-[(\Gamma/h)^2+1]^{-1}<0$. By switching on the coupling, $J>0$ the correlation function increases and for $J>J_c$ it becomes antiferromagnetic, $C>0$. 
Numerical results of the DMRG calculations are shown in the insets of Fig.\ref{fig:corr_func1} for $h=0$ (upper panel) and $h=0.4$ (lower panel).  
The curves can be scaled to a master curve by using the combination $C(J,L)L^{1/4}$ see Eq.(\ref{TIC_corr}) and $(J_c-J)L$ according to Eq.(\ref{TIC_gap}), which are shown in the main figures. For $h=0.4$ the correlation function goes below zero for sufficiently small value of $J$. To check this behaviour we magnified the correlation function near the transition point for $h=0.4,0.8$ and $1.2$, which is shown in Fig.\ref{fig:corr_func2}.
It is seen in this figure that for far enough to the transition point, $J_c-J>\Delta J(L)$, $C(J)$ has an approximately linear increase and there is no noticeable size-dependence. Critical fluctuations perturb this behaviour at $J_c-J \approx \Delta J(L) \sim 1/L$, when $C(J)$ starts to grow very fast. Extrapolating the linear behaviour up to $J_c$, which is calculated through the location of the maximum of the entanglement entropy in Sec.\ref{sec:entropy_max}, we obtain the limiting value of the correlation function:
\be
\lim_{J \to J_c^{-}}\lim_{L \to \infty} C(L,J,h)=C^{-}(h)<0\;,
\label{C_limit}
\ee
which is ferromagnetic. This limiting value is shown as a function of $h$ in the left inset of Fig.\ref{fig:corr_func2}. For small $h$ we notice a quadratic behaviour: $|C^{-}(h)| \approx 0.019 (h/\Gamma)^2$. Inspecting the slope of the linear part, $sl(h)$, we notice that its value is approximately proportional to $C^{-}(h)$. Their ratio is around $sl(h)/C^{-}(h)\sim -6.5$ and depends only weakly on $h$.
The perturbed part of the curves in Fig.\ref{fig:corr_func2} can be put to a master curve in terms of the scaling variables: $C(J,L)L^{1/4}$ vs. $(J_c-J)L$, which is shown in the right inset of this figure. As seen in the inset scaling works only up to a critical end-point.

We can conclude our numerical results in the following way. The correlation function in the thermodynamic limit is antiferromagnetic, $C(J)<0$ for $J>J_c$ which goes to zero as $C(J) \sim [J-J_c)^{1/4}$, both at $h=0$ and for $h>0$. Also the correlation length is divergent with the Ising exponent, $\nu=1$. For $h>0$ the correlation function is discontinuous at $J_c$, it has a jump $C^{-}(h)$, as defined in Eq.(\ref{C_limit}). Consequently the phase transition is mixed order in this case.

\subsection{End-to-end correlations}
\label{sec:end_end_corr}

\begin{figure}[h!]
\begin{center}
\includegraphics[width=8.6cm]{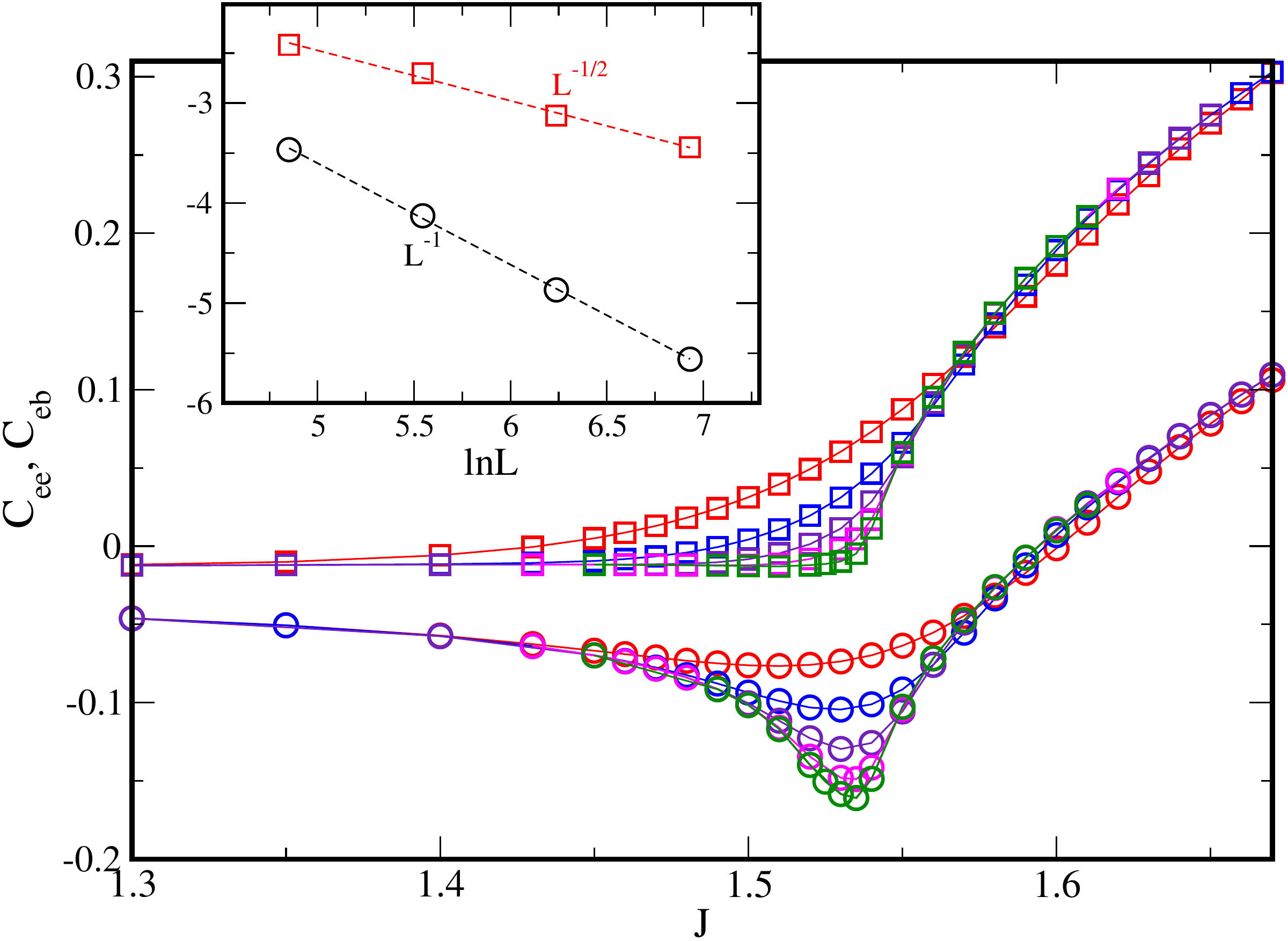}
\end{center}
\caption{\label{fig:corr_end_end}(Color online) End-to-end $C_{ee}$ (denoted by $\circ$) and end-to-bulk $C_{eb}$ correlation functions (denoted by $\square$) for $\Gamma=1.5$ and $h=0.4$ as a function of $J$ for different lengths of the open chain. $L=32$ $\circ$, $L=64$ \textcolor{red}{$\circ$}, $L=128$ \textcolor{blue}{$\circ$}, $L=256$ \textcolor{violet}{$\circ$}, $L=512$ \textcolor{magenta}{$\circ$} and $L=1024$ \textcolor{green}{$\circ$}. Inset: scaling of the finite-size correction to the minimum value of $[C_{ee}(J_c,L)+0.193]$ (denoted by \textcolor{red}{$\square$}) and the radius of curvature at the minimum, $R(J_c)$, (denoted by $\circ$), as a function of $L$ in log-log scale, see text.}
\end{figure}

We have also studied the behavior of the end-to-end correlation function, $C_{ee}=-\langle \sigma_1^z \sigma_{L}^z \rangle$, which is calculated for open chains. Here we should note that the end-spins are loosely connected to the chain. For example in the classical limit, $\Gamma=0$, $C_{ee}$ changes sign at $h=J$, so that for $1<h/J<2$ we have $C_{ee}=-1$, while the bulk correlations are $C=1$. To avoid such an independent surface transition we keep in the following $h<J$ even for $\Gamma>0$. As an example we consider $\Gamma=1.5$ and $h=0.4$ and plot $C_{ee}$ for different values of the antiferromagnetic coupling in Fig.\ref{fig:corr_end_end}. For a comparison we show in this figure also the end-to-bulk correlations, $C_{eb}=-\langle \sigma_1^z \sigma_{L/2}^z \rangle$, which is defined now with open boundary conditions.
It is seen in this figure that for both type of correlations a singularity develops around the phase-transition point at $J_c=1.52$. The end-to-end correlations show a minimum, and the value at the minimum, $C_{ee}(J_c,L)$, approaches an asymptotic value, $C_{ee}(J_c,L=\infty)\approx -0.193$ as a power: $\sim L^{-\eta^{'}_s}$ with an exponent $\eta^{'}_s\approx 0.5$. This is illustrated in the inset of Fig.\ref{fig:corr_end_end}, in which $C_{ee}(J_c,L)+0.193$ is plotted vs. $L$ in log-log scale. In this inset we have also studied the radius of curvature at the minimum, ${\cal R}(J)$, which is found to scale as ${\cal R}(J) \sim 1/L$. We can interpret ${\cal R}(J)$ as the distance from the critical point in a finite system, having a correlation length, $\xi \sim L$. Consequently at the transition point for the end-to-end correlations there is a diverging length, which scales with the Ising exponent, $\nu=1$ but at the same time its derivative is discontinuous in the thermodynamic limit.

For the end-to-bulk correlation function the analysis of the numerical data is more difficult. For the largest finite systems, $L=512$ and $1024$, we notice the appearance of a minimum near the phase-transition point and we expect a similar type of behaviour, as found for the end-to-end correlation function.

\section{Discussion}
\label{sec:disc}

In this paper we have re-examined the properties of the phase transition of the antiferromagnetic quantum Ising chain in the presence of a longitudinal field. Previous studies have focused on the excitation spectrum of the system and vanishing gaps are observed at the phase-transition line. Finite-size scaling of the low-energy excitations at the transition line are found the same behaviour for $h/\Gamma>0$ as for $h/\Gamma=0$, i.e. at the TIM point. Therefore it was generally accepted that the phase-transition for $h/\Gamma>0$ is controlled by the TIM fixed point. Here we have studied other physical quantities of the model, entanglement entropy and spin-spin correlation function, and have clarified this issue further.

First we have introduced and applied a quantum block RG method which is known to correctly describe the properties of the TIM fixed-point in the large block-size limit. Introducing a non-zero longitudinal field to the RG transformation the TIM fixed-point is found unstable, signalling the possibility that the phase-transition for $h/\Gamma>0$ belongs to a different universality class. Quantitative results about the properties of the phase transition have been calculated by the DMRG method. The location of the phase-transition line is obtained as the extrapolated value of the position of the maximum of the entanglement entropy. Finite-size scaling of the entanglement entropy at the transition line is found compatible with the behaviour at the TIM point, thus having a central charge $c=1/2$. The correlation length is extracted from the scaling of the entanglement entropy and for $J<J_c$ it is found to scale as in the TIM.

\begin{figure}[h!]
\begin{center}
\vskip5mm
\includegraphics[width=8.6cm,angle=0]{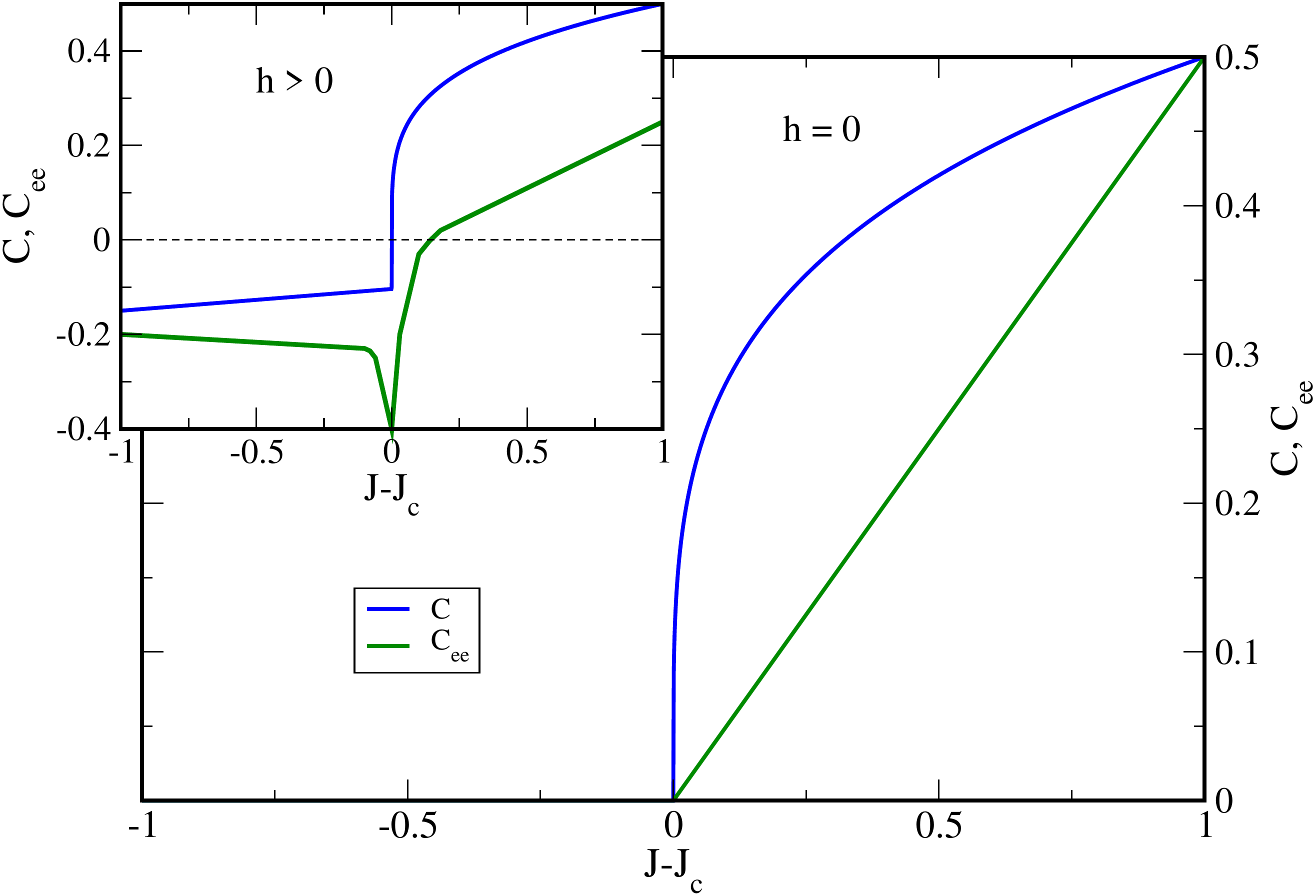}
\end{center}
\vskip-5mm
\caption{\label{fig:illustration}(Color online) Illustration of the coupling dependence of the spin-spin correlation function in the bulk ($C$) and between endspins ($C_{ee}$) at $h=0$ (main panel) and at $h>0$ (inset).}
\end{figure}

The AFM or FM order in the system is characterised by the spin-spin correlation function, what we have calculated for large periodic chains with different lengths. Starting from the AFM phase, $J>J_c$, the correlation function is found to vanish at the phase-transition point in the same form, as at the TIM, thus having the same critical exponents. Surprisingly, by passing the phase transition line, the correlation function shows a discontinuity with a limiting finite FM value. Repeating the same calculations for the end-to-end correlations in open chains a minimum is observed at the phase-transition point having a discontinuity in the derivative. The behaviour of the correlation functions around the phase-transition point is illustrated in Fig.\ref{fig:illustration} for $h=0$ and in the inset for $h>0$.

These findings can be interpreted that the phase transition in the antiferromagnetic quantum Ising chain in the presence of a longitudinal field is of mixed order. The length-scale, $\xi$, is divergent according to Eq.(\ref{TIC_gap}) at both sides of the transition point. Also the spin-spin correlation function shows AFM quasi-long-range order at the transition point having the TIM exponent below Eq.(\ref{TIC_corr}). But coming from the FM side the FM order vanishes discontinuously at the transition point. 

A similar type of discontinuity occurs near the CEP point which can be evaluated by two points of view. On the one hand, being in the classical limit, $J/\Gamma \to \infty$ the correlation function has a jump from $C=-1$ to $C=1$ at $h/J=2$ by increasing $J$. On the other hand, at the CEP point the ground state is infinitely degenerate, these are denoted by $|\psi_k^{(0)}\rangle$, with $k=1,2,\dots,K(L)$. The entropy per site is given by $\lim_{L \to \infty}\ln K(L)/L=\ln \tau$, with $\tau=(1+\sqrt(5))/2$ being the golden mean ratio\cite{hard_rods}. One can find a finite FM order by calculating the spin-spin correlation function in the low-temperature limit, when the degenerate states have equal weights, $|\Psi^{therm}\rangle=K(L)^{-1/2}\sum_k |\psi_k^{(0)}\rangle$. In the thermodynamic limit it is $C^{therm}_{CEP}=-0.2$. Switching on quantum fluctuations $1 \ll J/\Gamma < \infty$ the ground state becomes non-degenerate through an order through disorder phenomena\cite{villain} when the ground state is given as combination of the degenerate states with well defined weights: $|\Psi^{quant}\rangle=\sum_k c_k |\psi_k^{(0)}\rangle$. Also in this  limit the correlation function maintains finite FM value $C^{quant}_{CEP} < 0.$. Comparing its value with $C^{therm}_{CEP}$ we notice a jump, since $C^{quant}_{CEP}-C^{therm}_{CEP} \approx -0.16$.

Mixed-order transitions have already been observed in different systems: in  classical spin-chains with long-range interactions\cite{anderson1969exact,thouless1969long,dyson1971ising,cardy1981one,aizenman1988discontinuity,slurink1983roughening,bar2014mixed,angles}; in models of depinning transition\cite{PS1966,fisher1966effect,blossey1995diverging,fisher1984walks}; in percolation models with glass and jamming transition\cite{gross1985mean,toninelli2006jamming,toninelli2007toninelli,schwarz2006onset,liu2012core,liu2012extraordinary,zia2012extraordinary,tian2012nature,bizhani2012discontinuous,sheinman2014discontinuous} and in models at a multiple-junction or k-booklet geometry\cite{igloi91,cardy91,juhasz,grassberger}, for a recent review see in Ref.\cite{bar2014mixed1}. In our model the mixed-order transition is probably the consequence of competing AFM interactions and longitudinal fields in the presence of quantum fluctuations. In an analogy we might expect that mixed-order transition takes place in the multispin-version of the model\cite{penson} for $m=3$, or in the $q$-state FM quantum Potts-chain\cite{solyom} in the presence of a $q$-periodic longitudinal field. 

\begin{acknowledgments}
This work was supported by the National Research Fund under Grants No. K128989, No. K115959 and No. KKP-126749. The authors thank to J.-Ch. Anglés d'Auriac and H. Rieger for previous collaboration on this problem.
\end{acknowledgments}

\end{document}